\documentclass[aoas,dvips]{arximspdf}

\doi{10.1214/10-AOAS432}
\volume{4}
\issue{4}
\pubyear{2010}
\firstpage{1649}
\lastpage{1651}

\begin{document}
\begin{frontmatter}

\title{Remembering Leo}
\runtitle{Remembering Leo}

\begin{aug}
\author[A]{\fnms{Jerome H.} \snm{Friedman}\corref{}\ead[label=e1]{jhf@stanford.edu}}
\runauthor{J. R. Friedman}
\affiliation{Stanford University}
\address[A]{Department of Statistics\\
Stanford University\\
Stanford, California 94305 \\
USA\\
\printead{e1}} %adresu isvedimo komanda gale!
\end{aug}

% HISTORY:
\received{\smonth{10} \syear{2010}}

% ABSTRACT

% KEYWORDS

\end{frontmatter}

Leo Breiman was a unique character. There will not be another like him. I
consider it one of my great fortunes in life to have know and worked with him.
Along with John Tukey, Leo had the greatest influence on shaping my approach
to statistical problems. I did some of my best work collaborating with Leo,
but more importantly, we both had great fun doing it. I look back on those
years when we worked closely together with great fondness and regard them as
among the happiest and most fruitful of my professional career.

I first met Leo at an Interface conference at UCLA in 1974. He gave a talk on
nearest neighbor methods for classification. I had been working in
computational geometry using $k$-d trees to develop fast algorithms for finding
nearest neighbors. I mentioned this to Leo and he seemed quite interested. It
was clear even from that brief encounter that our interests coincided and that
we shared a common approach to statistical problems. At that time Leo was
working as a full-time statistical consultant in industry having resigned his
professorship at UCLA. After a brilliant career as a mathematical probabilist
he had totally changed his professional direction to applied statistics.

After that I had no contact with Leo for almost two years. In 1976 while
visiting CERN in Geneva I received a letter from Leo (there was no email then)
inviting me to give a talk at a conference he was organizing on
``Large and Complex Data Sets'' to be held in
Dallas in 1977. Although this topic was at the time far outside the mainstream of current statistical thinking, he was able to persuade (no doubt with
difficulty) the ASA and IMS to help sponsor it. Leo was a visionary. He
understood the need for what became known as data mining decades before the
name or discipline became fashionable. It was Leo's hope that the conference
would serve as a catalyst to move the statistical community in this direction,
at least a little. Although it took many years for this to happen, it was Leo
who started the process and he was a driving force for moving it forward throughout.

I got to know Leo better at the Dallas conference. One evening while I was
relaxing after having finished giving my talk, I saw Leo walking down a
hallway. He approached, handed me a stack of transparencies, and asked me to
present \emph{his} talk the next morning. He had to unexpectedly return to
Santa Monica to address a committee of the Democratic Party. Leo was a
candidate for the Santa Monica school board and needed their endorsement.
After some hesitation I agreed. Instead of seeing the sights of Dallas that
night with friends, I stayed in my room studying his transparencies\ until
late trying to understand the contents of his talk. That night Leo left for
Santa Monica and the next morning I gave his talk for him in Dallas. He
received the endorsement and was subsequently elected. I feel that in this
very small way I helped the school children of Santa Monica receive the
benefit of Leo's service. I also had the good fortune to meet Larry Rafsky at
the Dallas conference, leading to several years of collaboration on
graph-based methods for multivariate hypothesis testing.

My years of close collaboration with Leo began in the late 1970s when I
started working on the use of trees directly for classification rather than
just as a computation tool. Larry collaborated on part of this research. I
learned from Richard Olshen that Leo was working jointly with Chuck Stone on
similar approaches to tree-based methods as consultants for Technology Service
Corporation in Santa Monica. Some of the earliest and most important
developments on modern tree-based methods are contained in TSC technical
reports coauthored by Chuck and Leo from that period.

Richard told Chuck and Leo about our work on trees and Leo suggested that
Larry and I come to Santa Monica to discuss our respective approaches. It was
at that meeting that what became know as CART was initiated. Leo made the
suggestion that we form a collaboration to synthesize our approaches. He also
suggested that we write a monograph since, as he put it, ``No
statistics journal would ever publish this.'' Soon after,
Larry had to drop out due to the constraints of other commitments. Richard, who
had been doing wonderful work on the theoretical aspects of trees as well as
innovative applications to medical data, was invited to join. Thus Breiman,
Friedman, Olshen and Stone (BFOS) and CART were born. Shortly after that Leo
returned to academia, joining the Statistics Department at Berkeley.

From that time for almost ten years Leo and I worked closely together. The
routine was always the same. Every Thursday I would go up to Berkeley, arriving
at about noon. We would go to lunch and then back to Leo's office in Evans
Hall. There we would work without interruption (except for short breaks so
that I could feed my parking meter on Hearst Avenue) until around 5 pm or until
our progress began to wane. Leo would always decide when to end the working
session by saying, ``Jerry, let's go have a
beer.'' We would then adjourn to Shattuck Avenue Spats to
share a pitcher of beer followed by dinner at a good Berkeley restaurant. It
was then back to Stanford for me until next Thursday.

It was at Spats that the name for the ACE algorithm emerged. We were bouncing
around possible names over a pitcher when I suggested ACE as an acronym for
alternating conditional expectation. Leo didn't like the name and I did. We
both could be equally stubborn. Suddenly, almost out of nowhere Leo said,
``Jerry, you win. It's ACE.'' I was
surprised. Leo seldom gave in so suddenly. He said, ``Look over
there,'' pointing through the window to across Shattuck
Avenue. There was a huge sign with a big red ``ACE'' on it. It was an ACE Hardware store, and Leo felt it
was destiny.

In the late 1980s our research interests began to diverge and we no longer
closely collaborated. We did come together again in the early 1990s to work on
multiple response regression. Again it was great fun but by that time internet
infrastructure had improved to the point where it was possible to collaborate
effectively without being face-to-face, so my trips to Berkeley were fairly
infrequent. Leo wanted to call the procedure we developed ``Curds and Whey.'' I didn't like the name but since I had won
on ACE I gave in.

After that we more or less went our separate research ways, both pursuing
ensemble methods from different perspectives. Leo did his landmark work on
bagging that ultimately led to random forests, and I concentrated on boosting
approaches. But even when we were not collaborating Leo's influence on my work
was always present. All that I had learned from him over our years of
collaboration guided much of what I did, and still do. Leo had a special
approach to research that he taught me and no doubt his students as well. We
were all lucky to have known and worked with him.

I miss Leo and think of him often. Whenever I get a new idea that I think may
have merit, one of my first instincts is to call Leo and see what he thinks. Of
course, this is no longer possible. Leo has had a huge impact on statistical
science that has not, and no doubt will not, diminish with time. I was able to
learn first hand how he approached and solved problems and especially the joy,
enthusiasm, and passion he brought to his work. This has served me very well.

\printaddresses

\end{document}